%% file: main.tex
\pdfoutput=1
\documentclass[conference]{IEEEtran}
\IEEEoverridecommandlockouts
\usepackage{amsmath,amssymb,amsfonts}
\usepackage{algorithmic}
\usepackage{graphicx}
\usepackage{textcomp}
\usepackage{xcolor}
\usepackage{todonotes}
\usepackage{hyperref}
\def\BibTeX{{\rm B\kern-.05em{\sc i\kern-.025em b}\kern-.08em
    T\kern-.1667em\lower.7ex\hbox{E}\kern-.125emX}}
\usepackage[numbers]{natbib}
\usepackage{mdwtab}

\begin{document}

\title{Understanding Cache Boundness of ML Operators on ARM Processors}

\author{
\IEEEauthorblockN{Bernhard Klein}
\IEEEauthorblockA{\textit{Institute of Computer Engineering} \\
\textit{Heidelberg University, Germany}\\
bernhard.klein@ziti.uni-heidelberg.de}
\and
\IEEEauthorblockN{Christoph Gratl and Manfred Mücke}
\IEEEauthorblockA{\textit{Materials Center Leoben, Austria} \\
christoph.gratl@mcl.at \\
manfred.muecke@mcl.at
}
\and
\IEEEauthorblockN{Holger Fröning}
\IEEEauthorblockA{\textit{Institute of Computer Engineering} \\
\textit{Heidelberg University, Germany}\\
holger.froening@ziti.uni-heidelberg.de}
}

\maketitle

% include parts
\input{abstract.tex}
\input{introduction.tex}

\input{related_work.tex}

\input{methodical_setup.tex}
\input{performance_analysis_float.tex}

\input{performance_analysis_quantized.tex}
\input{summary.tex}
\input{acknowledgment.tex}
\bibliographystyle{IEEEtranN}
\bibliography{main}
\input{appendix.tex}

\end{document}

%% file: abstract.tex
% abstract
\begin{abstract}
Machine Learning (ML) compilers like TVM allow a fast and flexible deployment on embedded CPUs.
This enables the use of non-standard operators, which are common in ML compression techniques.
However, it is necessary to understand the limitations of typical compute-intense operators in ML workloads to design a proper solution.
This is the first in-detail analysis of dense and convolution operators, generated with TVM, that compares to the fundamental hardware limits of embedded ARM processors.
Thereby it explains the gap between computational peak performance, theoretical and measured, and real-world state-of-the-art results, created with TVM and openBLAS\footnote{\url{https://www.openblas.net}}.
Instead, one can see that single-precision general matrix multiply (GEMM) and convolutions are bound by L1-cache-read bandwidth.
Explorations of 8-bit and bit-serial quantized operators show that quantization can be used to achieve relevant speedups compared to cache-bound floating-point operators.
However, the performance of quantized operators highly depends on the interaction between data layout and bit packing.
\end{abstract}

% keywords

%% file: introduction.tex
\section{Introduction}

% Context
Although machine learning is meantime ubiquitously deployed, it is still nascent and developing fast.
One can observe a continuous stream of innovation, resulting in new methods, model architectures and use cases that benefit from ML's inherent property to model complex input-output relations with high accuracy and outstanding generalization.
Currently, this continuous development results in fast-changing basic operations of artificial neural networks.
One prominent example are Capsule networks, in which scalar-valued neurons are replaced by small matrices in order to capture more complex relationships.
There exist various reports on the inflexiblity of ML frameworks to address the needs of such innovation~\cite{stuckinarut}, as they are mainly designed for today's need, but obviously cannot anticipate tomorrow's innovation.
Due to the diminishing returns of CMOS scaling, the number of distinct processor architectures and technologies is rapidly growing, further increasing this problem as each platform requires a set of corresponding libraries.

The ubiquity of ML also requires the deployment on embedded and edge platforms, which are substantially limited in their resources, including processor, memory, network, and battery life, among others.
To address the gap between model requirement and hardware capability, a plethora of methods exist to compress the models.
Examples include quantization~\cite{roth2020}, which reverts from floating point values to low-bit-width alternatives for parameters such as weights and activations, pruning which introduces sparsity in the parameters to avoid computations~\cite{DBLP:journals/corr/abs-1810-05270}, and architecture search that basically trains not only parameters but also the hyperparameters that represent model architecture~\cite{DBLP:journals/corr/ZophL16}.
Notably, all techniques aim to maintain prediction accuracy while minimizing either the amount of computations, or parameters, or both.

% Gap
One of the most promising solutions to address the gap between ML innovation and hardware trends is code generation, which gears to automatically generate high-performance operators for a given target platform, thereby providing an alternative to hand-tuned, manually-written libraries.
One of the most prominent examples of code generation specialized for ML is TVM~\cite{tvm}, which is based on graph-level and operator-level optimizations and employs a learning-based cost model for automated optimization~\cite{chen2018autotuner}.
Experimental results have shown that the performance of generated operators is on-par with state-of-the-art, hand-tuned libraries for various target platforms~\cite{tvm}. 

Similarly to new operators resulting from ML innovation, compression techniques usually also require custom operators:
quantization can reach extremely low-precision representations, ultimately binarized values~\cite{rastegari2016}, while pruning needs to express the found sparsity in the operator, requiring for instance compressed-sparse row formats or masks that represent non-zero values.
Similarly, architecture search can result in complex-connected networks, which could also benefit from custom operators for efficiency reasons.
As a result, it seems promising to explore the capabilities of code generation for model compression, such that deployment is feasible with high efficiency, high productivity and performance comparable to hand-tuned libraries.

% Innovation
This work is mainly concerned with executing specialized ML operators on ARM processors as pervasively available embedded systems, analyzing performance limiters, and exploring countermeasures such as model compression. 
To be more concrete, we provide a detailed analysis of computationally intense operators, including convolutional and dense ones, on embedded ARM processors. 
As the main observation of this analysis is that such operations are bound by memory-cache bandwidth, we subsequently employ quantization as compression technique to reduce the necessary data volume.
We rely on code generation to solve the problem that the resulting specialized operators are often not supported by dedicated libraries.
For that, we select TVM due to its proven functionality, active community, and support for quantized computations~\cite{cowan2018,cowan2020}.
We furthermore choose bit-serial forms of computation on ARM processors, which allows for arbitrary reduced-precision representations on various platforms~\cite{umuroglu2019}.    
The bit serial approach does not scale according to reduced data size and moreover, it does not seem to be bound by cache bandwidth, at least not in regard to the cache-bound model discussed in this work. 
Also, as in our opinion quantization shall not be limited to certain data types, such as the de-facto industry standard of 8-bit integer, any quantization option has to be flexible for bestmost trading among accuracy and performance.

In particular, this work makes the following contributions:
\begin{enumerate}
\item 	Measure the computational peak performance and memory bandwidths for caches and RAM of selected ARM processors.
\item 	Benchmarking of convolutional and dense operations on embedded ARM processors, including auto-generated code and comparison with  openBLAS.
\item 	Detailed performance analysis to understand the disparity between sustained and theoretical peak performance, resulting in a cache-bound model.
\item 	Benchmarking 8-bit and bit-serial results, and discussing observed behavior based on the cache-bound model.
\end{enumerate}

%% file: related_work.tex
\section{Related Work}
% DL Compiler
Static inference libraries, like ARM's Compute Library\footnote{\url{https://developer.arm.com/ip-products/processors/machine-learning/compute-library}} or other specialized, hand-tuned, ultra-low-precision operators~\cite{han2020,schindler2017} achieve impressive performance, but make it difficult to combine multiple compression techniques if not provided by the library. However, the combination of such techniques, especially quantization and pruning, can be very effective~\cite{han2016deep}.

In contrast, deep learning compilers like TVM, Tensorflow's XLA~\cite{abadi2016tensorflow} and Lift~\cite{steuwer2017lift} close the gap between high-level machine learning frameworks and deployment on hardware in a flexible way. 
Their independence from static libraries enables straightforward research of new ML methods, the usage of non-standard operators, and allows to use compression techniques and to combine them with little effort. 

% TVM auto-tuning
Auto-tuning is one of the most important features of TVM. 
With an automated end-to-end feedback loop, executing runtime measurements, and a domain-specific machine-learning-based cost model, optimal parameters can be found~\cite{chen2018autotuner}. 
One step further, it is possible to exploit AutoTVM and use it in the decision process of hardware design~\cite{diamantopoulos2020agile}.

% TVM & quantization
While import and execution of previously quantized 8-bit models is supported using a specialized QNN dialect~\cite{jain2020qnn}, ultra-low-precision quantization is still an open research topic. 
Whereby, \citeauthor{cowan2018}~\cite{cowan2018} proved that such low-precision operations are feasible with TVM and, using a hardware-aware bit-serial algorithm, a relevant speedup over the default floating-point implementation can be achieved. 
Moreover, TVM's scheduling mechanism can be used for high-level optimizations and program synthesis to find highly optimized low-level operators~\cite{cowan2020}.

Our work does not focus on a new technique reducing the inference time even further.
It focuses on the analysis and understanding of performance bounds of compute-intense ML operators and compare this with the fundamental hardware limits.
It brings memory access as the limiting factor into mind and thus motivates ML compression techniques like low-precision representation, which can reduce the memory pressure, and thus the overall processing latency.

%% file: methodical_setup.tex
\section{Methodological Setup and Performance Baseline}

\subsection{Auto-tuning Methodology}
The interface of AutoTVM is designed to tune neural networks which are represented in Relay~\cite{roesch2018relay}, TVM's high-level IR. However, in our work, specific single operators are evaluated. Creating many single-layer neural networks, containing exactly the operations to be examined, auto-tune them and save the tuned parameters to a logfile, enables reuse and thus auto-tuned operator execution in the manual examination mode.

For regular data types, float32, unsigned and signed int8, the XGBTuner with its xgboost cost model is used~\cite{chen2016xgboost}. 
However, this tuner is---because of a not yet fixed issue---not compatible with some of the bit-serial operators, therefore, all bit-serial operators are auto-tuned with the random tuner. 
In principle the tuner can have a relevant impact on the tuned parameters and thereby on the final inference time.
However, for bit-serial dense and convolution operators, the search space is highly restricted due to the bit-packing implementation resulting in less freedom in the parameter selection. 
Therefore, the impact of auto-tuning is relative small and the selection of the tuner less critical.

TVM release 0.7\footnote{TVM: release v0.7 commit efdac9439506d1de5eec91ecc795982c78e41909} is used and compiled with openBLAS support.
	
\subsection{Target Architecture}

Our experiments are performed on ARM Cortex-A53\footnote{\url{https://developer.arm.com/ip-products/processors/cortex-a/cortex-a53}}, Broadcom BCM2837 and ARM Cortex-A72\footnote{\url{https://developer.arm.com/ip-products/processors/cortex-a/cortex-a72}}, BCM2711.

\subsubsection{Peak Performance}
the computational theoretical peak performance,
\begin{equation}
	p_{peak} = frequency \cdot cores \cdot \frac{FLOP}{instr} \cdot \frac{instr}{cycle} \cdot SIMD_{width}
\end{equation}
represents directly the maximum possible performance, assuming that all compute resources can be fully utilized.
Considering multiply-accumulate operations (MACs) with 2\,FLOPs per instruction, one NEON MAC per cycle and a NEON SIMD width of $128$\,bit, this leads to a single-precision peak performance of $38.4$\,GFLOP/s and $48.0$\,GFLOP/s for Cortex A53\,(1.2\,GHz) and A72\,(1.5\,GHz), respectively.

In order to verify theoretical expectations, a small benchmark program was written that executes many MACs, NEON's \textit{VMLA} instructions making only use of on-register operations, avoiding any other memory access. 
For a fair comparison, the total amount of MACs in a GEMM is distributed to all cores and multi-threading effects are included in the measurement, which is plainly visible for small matrices.
The critical part is written in assembly to ensure that no compiler optimizations distort the workload.
The benchmark code is publicly available\footnote{\url{https://github.com/UniHD-CEG/arm-peak}}.

The measured performance, shown in Table~\ref{tab_gemm_fp32_performance_pi3} and \ref{tab_gemm_fp32_performance_pi4}, confirms the compute boundary, provided that the workload is large enough to hide the overhead of multi-threading.
Additionally, it confirms that one \textit{VMLA} instruction can be computed per cycle.

\subsubsection{Memory Bandwidth} 
 With the benchmark tool RAMspeed\footnote{\url{https://github.com/cruvolo/ramspeed-smp}} the read and write bandwidth with different block sizes is measured. 
Since the L1 data cache is 16\,KB for the ARM Cortex-A53 and 32\,KB for the Cortex-A72, a block size of 4\,KB is used to measure L1 cache bandwidth. 
To fit into the L2 caches---512\,KB Cortex A53 and 1\,MB Cortex A72---256\,KB sized blocks are used. 
Last, to measure the RAM bandwidth, a block size of 16\,MB is large enough to keep caching effects reasonably small. 
For the measurements shown in Tables \ref{tab_bandwidths_pi3} and \ref{tab_bandwidths_pi4}, RAMspeed is used in multi-threading mode with 4 threads to utilize all cores and with 1\,GB and 8\,GB of data per pass, respectively, the maximum that still fits into main memory.

\begin{table}[!t] \renewcommand{\arraystretch}{1.3} \caption{Measured Memory Bandwidth for ARM Cortex A53} \label{tab_bandwidths_pi3}
\centering
\begin{tabular}{c|c|r|r} 
\bfseries Memory  	& \bfseries Block Size & \bfseries Read BW & \bfseries Write BW \\ 
\hline
RAM 		& 16\,MB  	&  2040\,MiB/s 	&  1600\,MiB/s\\
L2 Cache 	& 256\,KB 	&  7039\,MiB/s 	&  3467\,MiB/s\\
L1 Cache 	& 4\,KB  	& 14363\,MiB/s 	& 23703\,MiB/s\\
\end{tabular}
\end{table}
\begin{table}[!t] \renewcommand{\arraystretch}{1.3} \caption{Measured Memory Bandwidth for ARM Cortex A72} \label{tab_bandwidths_pi4}
\centering
\begin{tabular}{c|c|r|r}
\bfseries Memory  	& \bfseries Block Size & \bfseries Read BW & \bfseries Write BW \\ 
\hline
RAM      & 16\,MB   &   3661\,MiB/s  &  2984\,MiB/s \\
L2 Cache & 256\,KB  &  12934\,MiB/s  &  7407\,MiB/s \\
L1 Cache & 4\,KB    &  45733\,MiB/s  & 30423\,MiB/s \\
\end{tabular}
\end{table}

\subsection{Operators}

\subsubsection{General Matrix Multiply}
in dense layers all neurons are connected with all neurons of the previous layer. 
They are a key component of classical ML models and can be represented as dense GEMMs with a non-linear activation function afterwards.

To compute a GEMM with squared matrices, based on three nested loops, approximately $N^3$ MACs are necessary, thus the total amount of arithmetic operations is $2 \cdot N^3$.
With an execution time $t$, the performance of a GEMM can be computed as:
\begin{equation}
p= \frac{2 \cdot MACs}{t} = \frac{2\cdot N^3}{t}
\end{equation}

%------------------------------------------------------------------------------%
\subsubsection{Convolutions}
Convolutional Neural Networks (CNNs) are not only state-of-the-art in computer vision, moreover, they spread in many ML domains as a general, very effective concept.
Thus, convolution layers are the heart of most modern ML models, and are the most resource and time-consuming part, too.  

It is a common practice to employ GEMMs to compute convolution operators, relying on IM2COL~\cite{chellapilla2006im2col}, but native convolution algorithms are also common.

ResNet~\cite{he2016resnet} is one of the most successful and wide-spread architecture and smaller variants are common on embedded devices.
The properties of all convolution layers of ResNet-18 are shown in Table~\ref{tab_resnet18_convolution_layers}. 
As in previous work, the first layer is excluded since the input layer is particularly sensitive to quantization and the input channel depth is too low for efficient bit packing~\cite{cowan2018}.

\begin{table}[!t] \renewcommand{\arraystretch}{1.3} \caption{Resnet-18 Convolution Layers} \label{tab_resnet18_convolution_layers}
\centering
\begin{tabular}{c|c|c|c|c|c|c|c|c|r}
\bfseries Name & $b$ & $c_{in}$ & $c_{out}$ & $h_{in}$ & $w_{in}$ & $k$ & $s$ & $p$ & MACs \\ \hline
C2 &1 &  64 &  64 &  56 &  56 & 3 & 1 & 1 & 124,010,496\\
C3 &1&  64 & 128 &  56 &  56 & 3 & 2 & 1 &  62,005,248\\ 
C4 &1&  64 & 128 &  56 &  56 & 1 & 2 & 0 &   6,422,528\\ 
C5 &1& 128 & 128 &  28 &  28 & 3 & 1 & 1 & 132,710,400\\ 
C6 &1& 128 & 256 &  28 &  28 & 3 & 2 & 1 &  66,355,200\\ 
C7 &1& 128 & 256 &  28 &  28 & 1 & 2 & 0 &   6,422,528\\
C8 &1& 256 & 256 &  14 &  14 & 3 & 1 & 1 & 150,994,944\\
C9 &1& 256 & 512 &  14 &  14 & 3 & 2 & 1 &  75,497,472\\
C10 &1& 256 & 512 &  14 &  14 & 1 & 2 & 0 &   6,422,528\\
C11 &1& 512 & 512 &   7 &   7 & 3 & 1 & 1 & 191,102,976\\
\end{tabular}
\end{table}

Like the GEMM, convolutions highly rely on MAC operations. 
For a batch-size $b$, pad $p$, stride $s$, input and output channels $c_{in}, c_{out}$, input and output image size $h_{in}, w_{in}, h_{out}, w_{out}$, and convolution kernel $\vec{k}$, 
the number of MACs in a convolution layer is:
\begin{align}
h_{out} &= \frac{h_{in} + 2 p}{s};   
w_{out} = \frac{w_{in} + 2 p}{s} \\
MACs &= b \cdot h_{out} \cdot w_{out} \cdot c_{in} \cdot c_{out} \cdot k_x \cdot k_y
\end{align}
Table \ref{tab_resnet18_convolution_layers} demonstrates the amount of diversity among layers with regard to computational complexity.
Notably, $3 \times 3$ convolutions are most compute intensive, and layers with a non-unit stride can lead to complex memory access patterns.

%% file: performance_analysis_float.tex
\section{In-Detail Performance Analysis}

\subsection{Matrix Multiply Performance}

Table~\ref{tab_gemm_fp32_performance_pi3} and \ref{tab_gemm_fp32_performance_pi4} illustrate the performance for various squared matrices, with results generated by TVM with and without auto-tuned parameters, and with external openBLAS library calls.
For all measurements, the overhead of multi-threading is dominating for small matrices.
However, for all matrices, the auto-tuned solution clearly outperforms the one without tuning, which has to fall back to default parameters, and even outperforms the hand-tuned BLAS library.

\begin{table}[!t] \renewcommand{\arraystretch}{1.3} \caption{GEMM Performance float32 Cortex A53} \label{tab_gemm_fp32_performance_pi3}
\centering
\begin{tabular}{c|c|c|c|c|c}
\bfseries N & \bfseries openBLAS &  \multicolumn{2}{c|}{\bfseries TVM}  & \multicolumn{2}{c}{\bfseries compute peak perf.}\\\hline
            &                    & \bfseries naive & \bfseries tuned & \bfseries measured &\bfseries theoretical \\ \hline 
            & \multicolumn{5}{c}{\bfseries \,GFLOP/s} \\ \hline
32   & 1.07 & 1.16  & 4.43  &    16.49 & 38.4 \\
128  & 4.96 & 2.07  & 6.58  &    37.38 & 38.4 \\
256  & 4.71 & 1.83  & 6.93  &    38.04 & 38.4 \\
512  & 4.87 & 0.60  & 5.06  &    38.15 & 38.4 \\
1024 & 4.99 & 0.54  & 5.01  &    38.18 & 38.4 \\
\end{tabular}
\end{table}

\begin{table}[!t] \renewcommand{\arraystretch}{1.3} \caption{GEMM Performance float32 Cortex A72} \label{tab_gemm_fp32_performance_pi4}
\centering
\begin{tabular}{c|c|c|c|c|c}
\bfseries N & \bfseries openBLAS &  \multicolumn{2}{c|}{\bfseries TVM}  & \multicolumn{2}{c}{\bfseries compute peak perf.}\\\hline
            &                    &  \bfseries naive & \bfseries tuned   & \bfseries measured &\bfseries theoretical \\ \hline 
& \multicolumn{5}{c}{\bfseries \,GFLOP/s} \\ \hline
32    &  3.01  &   3.59  &   9.20   & 21.92 & 48.0 \\
128   & 14.22  &   4.68  &  16.72   & 47.11 & 48.0 \\
256   & 14.86  &   4.77  &  17.24   & 47.83 & 48.0 \\
512   & 14.33  &   2.04  &  17.99   & 47.92 & 48.0 \\
1024  & 14.98  &   1.36  &  15.75   & 47.93 & 48.0 \\ 
\end{tabular}
\end{table}

The performance saturates for larger matrices, albeit, it is significantly lower than the expected peak performance of the CPU.

\subsection{Cache-bound Model}
In the best case scenario all computations are executed on fast local registers. 
However, there are obviously not enough registers for all matrix elements. 
In a simplified model, it can be assumed that for a vectorized MAC operation, the first operand can be kept in registers, so that the result is also accumulated in registers, but at least the second operand has to be read from some kind of memory.
Therefore, in the following discussion it is assumed, that for a MAC operation at least one operand has to be read from cache or RAM. 
Certainly this is a very simplified model since the variables, kept in registers, need to be exchanged after multiplied with all their counterparts, and the accumulated results have to be written back to memory, too. 
However, this is the most common operation and should dominate overall execution time.

\begin{figure}[!t]
\centering
\includegraphics[width=3.5in]{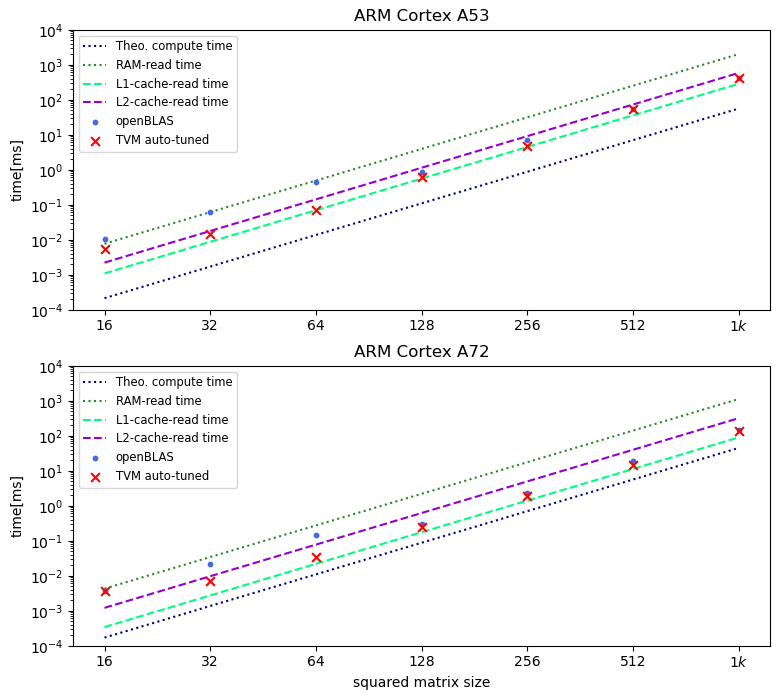}
\caption{Execution time over squared matrix size for general matrix multiply in double logarithmic representation. Together with hardware boundaries, theoretical compute time, time to read or write data to RAM or cache-memory.}
\label{fig_gemm_cache_bound}
\end{figure}

Fig.~\ref{fig_gemm_cache_bound} shows in double logarithmic scale the execution time of auto-tuned TVM and openBLAS with respect to the matrix size. Additionally the compute time representing the theoretical peak compute performance, and the time to read $4 \cdot N^3\,$bytes for float32 data type from L1, L2 cache and RAM memory are plotted. 
The bandwidth of the different memory types are listed in Tables \ref{tab_bandwidths_pi3} and \ref{tab_bandwidths_pi4}. 

For small matrix sizes, the execution times are somewhere in the range between RAM and cache bandwidth bounds, but the measured time is in the range of only a few microseconds, thus vulnerable to small systematic errors, like multi-threading overhead. 
However, the time for matrices with $N \geq 100$ strongly correlates with the L1 cache boundary. 
This suggests that the L1-cache-read bandwidth is not fast enough to keep the floating-point units fully utilized.
Therefore, the theoretical peak performance cannot be reached, and 
it becomes apparent that single-precision GEMM for ARM Cortex A53 and A72 processors is not compute-bound, but cache-bound.  

\subsection{Performance of Convolutions}

For the experiments, the ARM-specific \textit{conv2d spatial pack} operator with NCHW data layout~\cite{cuDNN} is used. There is also a NHWC~\cite{cuDNN} version available, but it performs much worse without auto-tuning and AutoTVM cannot be applied for this operator, thus this analysis is postponed until a fair comparison is possible.

\begin{figure}[!t]
\centering
\includegraphics[width=3.5in]{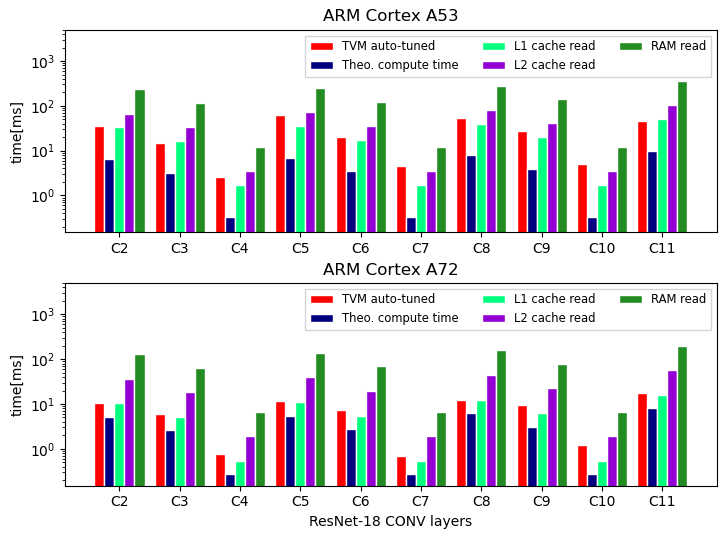}
\caption{Comparison of TVM's execution time of convolution layers of ResNet-18, to minimal theoretical compute and memory-read times. Mostly execution time correlates with L1 or L2 cache read times.}
\label{fig_conv2d_fp32_cache_bound}
\end{figure}

Fig.~\ref{fig_conv2d_fp32_cache_bound} sets the measured execution time for these convolutional operations in relation to the expected time for compute and memory read, again assuming that $4 \cdot $MAC\,bytes of data need to be read. 
In particular, one can see that the execution time for \textit{TVM auto-tuned} correlates with \textit{L1 cache read time}, though there are some notable exceptions.
Again, this execution time is far from \textit{theoretical compute time} as well as \textit{RAM read time}.

\begin{figure}[!t]
\centering
\includegraphics[width=3.5in]{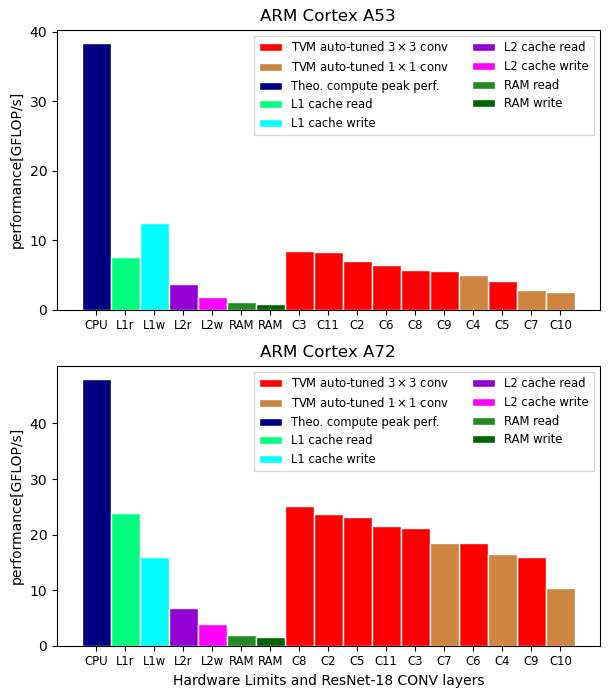}
\caption{Performance of ResNet-18 layers compared with theoretical compute peak performance and memory bandwidth limited performances. Layers are sorted in descending performance order.}
\label{fig_conv2d_fp32_perf}
\end{figure}

The performance in terms of GFLOP/s as shown in Fig.~\ref{fig_conv2d_fp32_perf} confirms this finding. 
Moreover, some of the compute intensive $3 \times 3$ convolutions can reach a little better performance than the L1-memory-bandwidth suggests. Probably this kernel layout allows optimizations regarding in-register reuse. However, this effect is not large enough to change the mainly dominating cache-bound behavior.

%% file: performance_analysis_quantized.tex
%\section{Abstract Performance Modeling for Quantized Operators}
\section{Performance Analysis of Quantized Operators}

The main take-away from the last section is that indepenent of the operator type, either GEMM or CONV, execution time is bound not by computational peak performance but memory access time, usually those of L1 caches.
While this is a surprising finding for HPC settings, apparently embedded processors have different design objectives.

Also, this cache boundness suggests that performance scaling can benefit most from reduced memory pressure.
Therefore, in the following we apply model quantization to reduce the size of operands, thereby also reducing the data volume fetched for a given operation.

\subsection{Measurements}
TVM supports bit-serial quantized convolution and dense operators for ARM CPUs for bipolar $(-1,1)^{b}$ and unipolar $(0,1)^{b}$ bit encodings for various bit widths~\cite{cowan2020}.
Whereas the unipolar variant is the more advanced quantization scheme, generally achieving better accuracies, it needs one additional subtraction and popcount instruction and is thus a little slower.
While the precision dimension is calculated sequentially, bit-packing along the spatial dimensions allows the usage of vectorized instructions~\cite{umuroglu2018bismo}.

The operators allow to set activation and weight bits independent from each other. It is common to use activations and weights with the same precision, or with larger activation precision than weight precision~\cite{roth2020}.
Still, in this work activations and weights are quantized equally.

For bit-serial operations, data has to be in a specific packed data layout. The weights can be pre-packed and thus do not need to be packed during runtime, but the activations require bit-packing just before the calculation. 

\subsection{Quantized GEMM}
\begin{figure}[!t]
\centering
\includegraphics[width=3.5in]{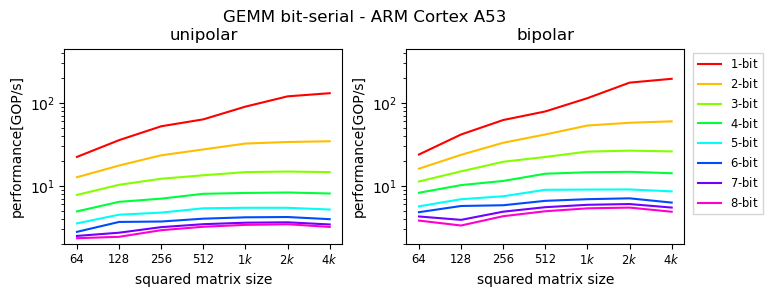}
\includegraphics[width=3.5in]{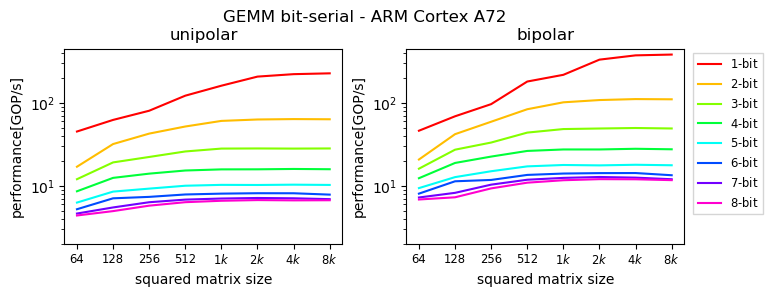}
\caption{Performance of bit-serial GEMM for different squared matrices, with the size of a dimension reported on the x axis. Lower bit widths achieve their maximum performance only with larger matrices.}
\label{fig_gemm_bitserial_performance}
\end{figure}

Fig.~\ref{fig_gemm_bitserial_performance} shows the performance of bit-serial GEMM for different matrix sizes. It is remarkable that for lower bit widths, even larger matrices are necessary to achieve maximum performance. For the extreme binary case it might not even reach its peak with $8k$ matrices, which is way larger than a typical dense layer in most ML workloads. 

To test whether bit-serial GEMM is also cache-bound, the required bandwidth to achieve measured performance is calculated.  
Depending on the used data type, it is assumed that per MAC $m$ one data read with $d$ bytes is necessary to sustain performance $p$ with execution time $t$. 
Thus, the required bandwidth $bw_{req}$ is:
\begin{equation}
bw_{req} = \frac{m \cdot d}{t} = \frac{p\cdot d}{2}
\end{equation}

\begin{figure}[!t]
\centering
\includegraphics[width=3.5in]{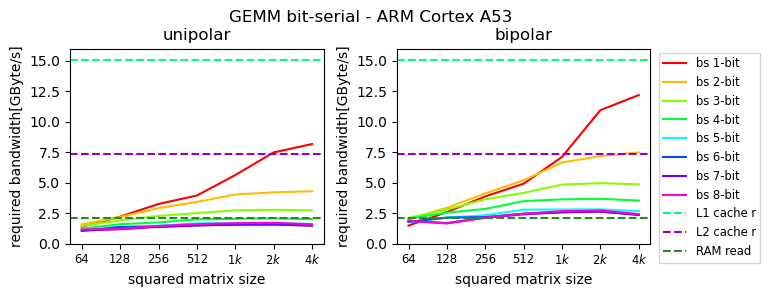}
\includegraphics[width=3.5in]{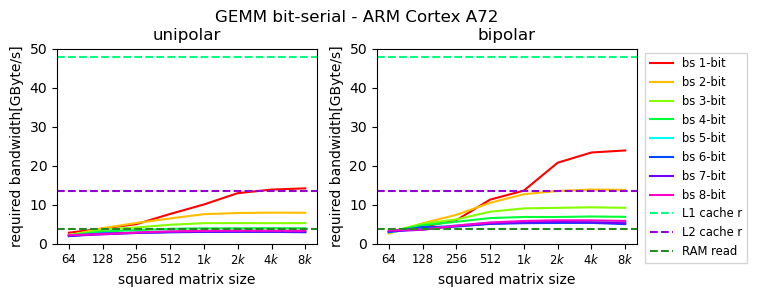}
\caption{Required bandwidth to reach measured performance according to cache-bound model for GEMM bit-serial operation.}
\label{fig_gemm_bitserial_cache}
\end{figure}

In Fig.~\ref{fig_gemm_bitserial_cache} this required bandwidth is plotted. 
With increased performances for larger matrices the requirements due to bandwidth increase too, but they all stay below the L1-cache-read bandwidth. 
While this suggests that bit-serial GEMM is not cache-bound, 
it has to be mentioned that other effects might be present in this case. 
One example is the efficiency of accessing bit-packed values. 
It is possible that the same packed data have to be read more than once per MAC, thereby the required bandwidth would also increase.
Whether bit-packed data increases the required cache reads per MAC is an open question that is left for future research as of now.
Furthermore, the mandatory bit-packing step for activations before the GEMM needs additional memory accesses, which are also not covered by the one-read-per-MAC assumption.

\subsection{Quantized Convolution}
\label{ch_quantized_convolution}

\begin{figure}[!t]
\centering
\includegraphics[width=3.5in]{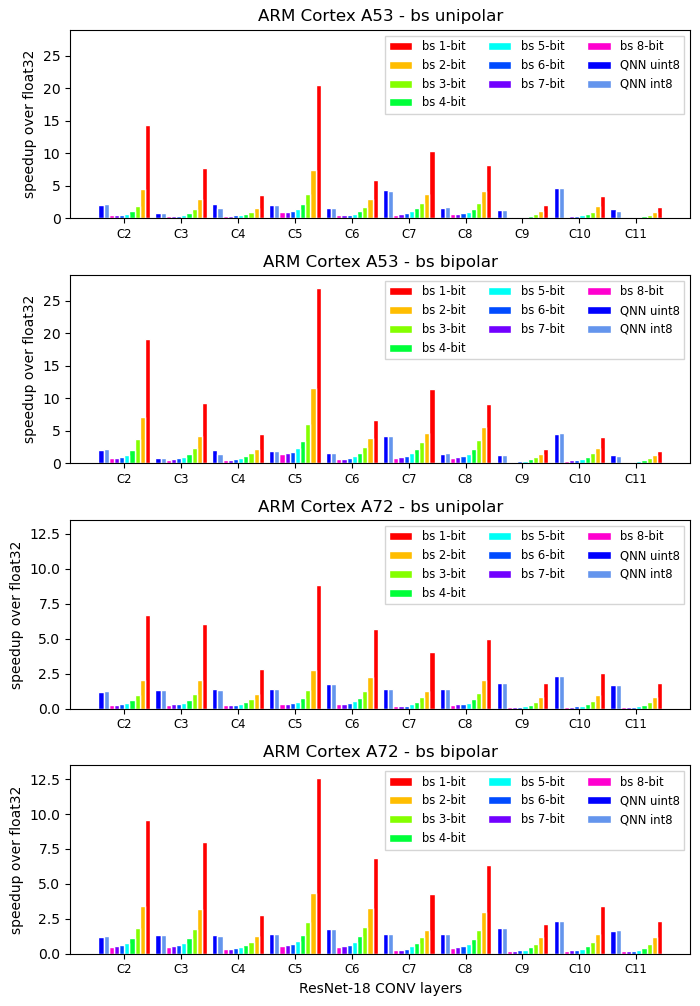}
\caption{Speedup over float32 for QNN 8-bit and bit-serial convolution for ResNet-18 layers.}
\label{fig_conv2d_bitserial_speedup}
\end{figure}

The speedup of the quantized ResNet-18 convolution layers is illustrated in Fig.~\ref{fig_conv2d_bitserial_speedup}. 
The performance of 8-bit QNN or bit-serial operators highly depend on the interaction between operator data layout and bit-packing format.
In general, $3\times 3$ convolutions benefit from higher computational complexity and data reuse, and therefore perform better than their $1 \times 1$ counterpart. 
A non-unit stride can lead to less efficient memory access especially for packed data. Furthermore, the bit-serial convolution operator uses NHWC data layout, which together with bit-packing leads to a performance decrease for small input sizes. 
For instance, layer 11 performs badly for the bit-serial operator, even though this operation has the highest MAC count. 
In contrast, 8-bit QNN with its NCHW layout is less sensible to the input size and it further seems to be more robust against $1 \times 1$ convolutions for smaller images than the floating-point baseline.
For the bit-serial approach, the computational complexity scales quadratically with the bit width, thus, the speedup of lower bit-width representations is significantly higher.

\begin{figure}[!t]
\centering
\includegraphics[width=3.5in]{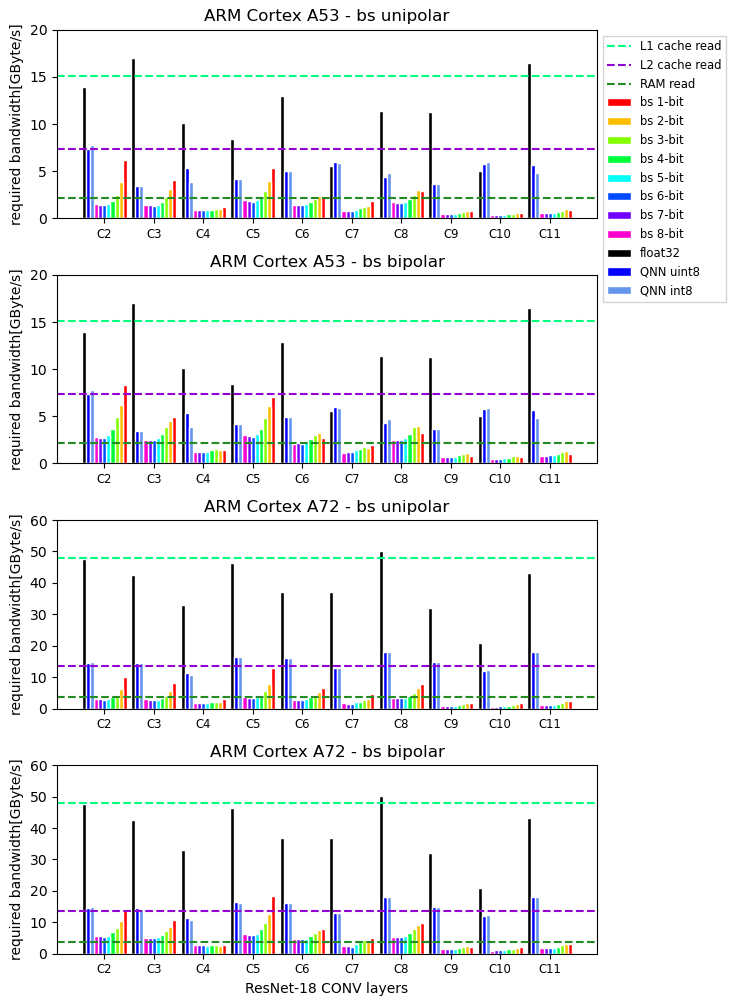}
\caption{Required bandwidth to sustain measured performance, according to cache-bound model of convolution operators. Comparison of float32, QNN 8-bit and bit serial to memory bandwidths.}
\label{fig_conv2d_bitserial_required_bandwidth}
\end{figure}

Fig.~\ref{fig_conv2d_bitserial_required_bandwidth} shows the required bandwidth to sustain measured performance.
The bandwidth requirements scale linear with the bit width, but the computational complexity quadratically, therefore, higher required bandwidths, due to the better overall performance, are observable for lower bit widths. 

More importantly, the required bandwidth indicates that, like for the GEMM, the L1-cache bandwidth is sufficient to provide data if only one read per MAC is necessary. Therefore, the 8-bit QNN and bit-serial convolutions are apparently not cache-bound.
However, multiple reads due to packed data and the overhead for bit packing are not covered in this model, but could be relevant. 

%% file: summary.tex
\section{Summary and Outlook}
First the fundamental hardware limits, the bandwidth of caches and RAM, and the computational peak performance are measured, whereby the measured peak performance fits well to the theoretical maximum.
The execution time of GEMM and convolution operators, generated with TVM and openBLAS, are compared against these hardware limits. 
The huge difference between computational peak performance and sustained performance of representative operators can be explained with a cache-bound model, which assumes that per MAC operation one memory read from cache, typically L1 cache, is necessary.

The performance of single-precision floating point operators correlates mainly with the limit created by the L1-cache-read bandwidth.  
To overcome this bottleneck, low-bit-width operators are explored: 8-bit QNN and the bit-serial approach for bit widths between 1 and 8 bit.
Both achieve relevant speedups compared to the floating-point baseline. 
The analysis of the required bandwidth to read bit-packed data and sustain measured performance indicates that quantized operators are not limited by cache-read bandwidth, at least not with the simple cache-bound model discussed in this work.
However, it becomes apparent that due to the bit-packed data structure, the operator data layout is crucial.
Moreover, to reach maximal performance of bit-serial GEMMs very large matrices, and thus a more efficient access to bit-packed data, is essential.

Relevant future directions include understanding the overhead of bit packing and access to packed data, scaling of memory accesses with problem size, and a corresponding refinement of the cache-bound model.
Such a model, would be of great benefit for the design of new operators and could also be included in an improved auto-tuning.

Ultimately, the study of operators with differently quantized activations and weights would be of great interest, especially from the point of view that bit packing is only necessary for activations, but packed data access applies for both.

In summary, we believe that the best choice in terms of quantization for a given ARM processor requires a detailed understanding of the underlying effects, so that methodological approaches can be pursued for corresponding optimizations, instead of brute-forcing the problem by exhaustive searches based on a vast number of individual performance experiments.

%% file: acknowledgment.tex
\section*{Acknowledgment}
The authors would like to thank Stephan Diestelhorst from Xilinx DCG for sharing his valuable experiences, which finally leads us to the cache-bound model idea.

Furthermore, we gratefully acknowledge the financial support under the scope of the COMET program within the K2 Center “Integrated Computational Material, Process and Product Engineering (IC-MPPE)” (Project No 859480). This program is supported by the Austrian Federal Ministries for Transport, Innovation and Technology (BMVIT) and for Digital and Economic Affairs (BMDW), represented by the Austrian research funding association (FFG), and the federal states of Styria, Upper Austria and Tyrol.

%% file: appendix.tex
\section{Appendix}

\begin{figure}[t]
\centering
\includegraphics[width=3.2in]{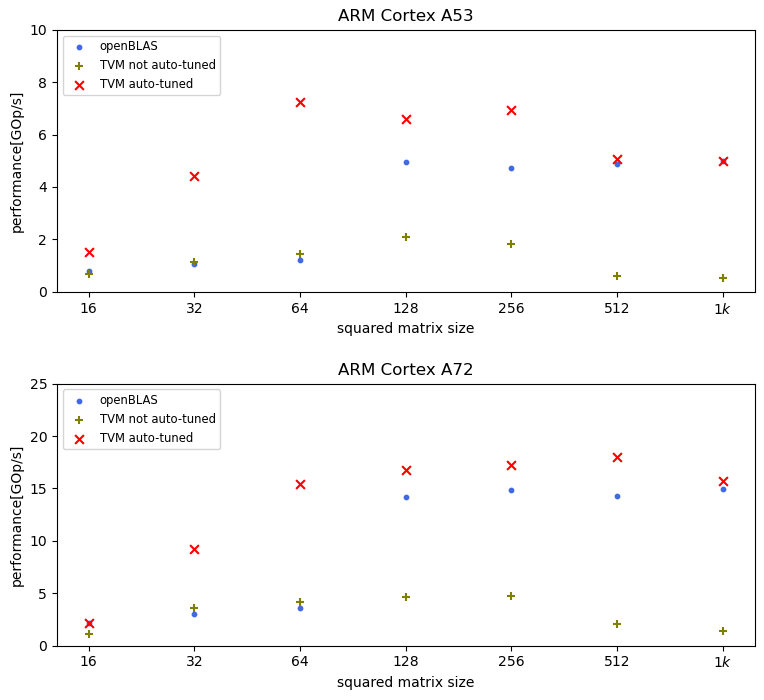}
\caption{Performance over matrix size for general matrix multiply. TVM with auto-tuned parameters outperforms TVM with not tuned parameters and openBLAS.}
\label{fig_gemm_performance}
\end{figure}

Fig.~\ref{fig_gemm_performance} shows the performance in GFLOP/s of floating point GEMM for TVM with and without auto-tuned parameters in comparison to openBLAS.
Auto-tuned TVM clearly outperforms the naive version and, moreover, also outperforms openBLAS.
For small matrices, systematic errors---like multi-threading overhead---are clearly visible.
In the middle regime auto-tuned TVM slightly outperforms openBLAS, probably due to parameters which are optimized to this specific hardware and matrix size.
For larger matrices auto-tuned TVM and openBLAS achieve comparable results.
It becomes apparent that hand-tuned general BLAS concepts are on-par with hardware and operator specific code generation.

However, it is clearly visible that all measured performances are far away from the theoretical peak performance, which can be explained with the cache-bound model.

\begin{figure}[b!]
\centering
\includegraphics[width=3.5in]{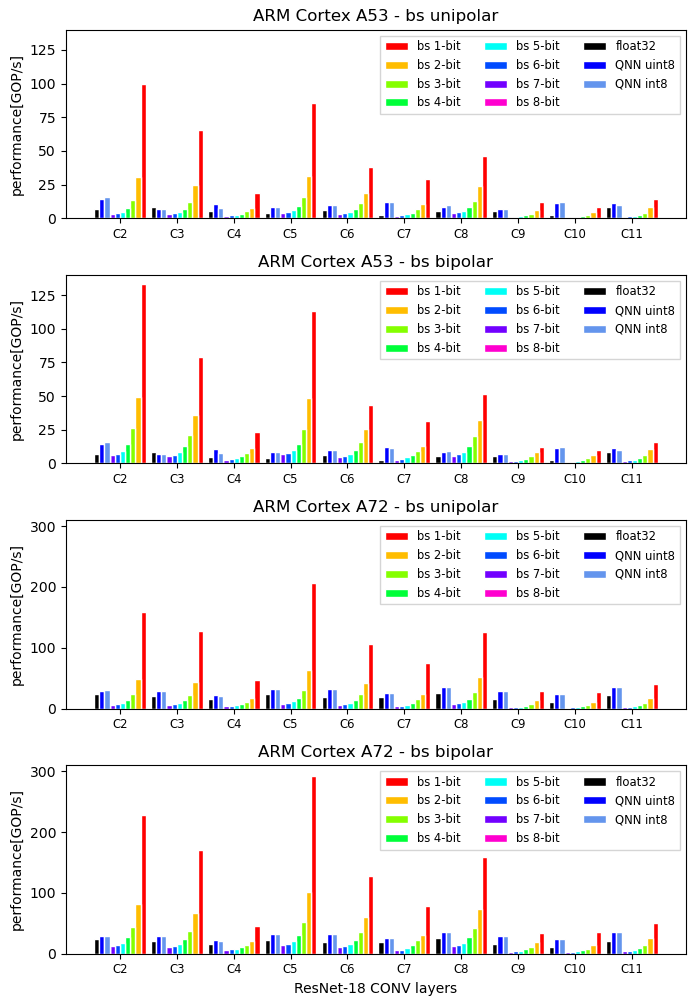}
\caption{Performance of bit-serial convolution operators of ResNet-18 layers in comparison to float32 and 8-bit QNN generated and auto-tuned with TVM.}
\label{fig_conv2d_bitserial_perf}
\end{figure}

Fig.~\ref{fig_conv2d_bitserial_perf} compares the performance of floating point with the quantized 8-bit QNN and bit-serial approaches for ResNet-18 convolution layers.
The performance of the different approaches for the diverse operators is highly related to the underlying data layout structure, as discussed in chapter ~\ref{ch_quantized_convolution}.
The computational complexity scales quadratically with the bit width for bit-serial forms, and thus larger bit widths perform worse than 8-bit QNN or even floating point. 
However, for smaller bit widths, this complexity scaling turns into a benefit, and lower bit-width representations significantly outperform floating point and 8-bit baselines.
Due to the slightly reduced computational complexity bipolar bit-serial forms achieve higher overall performances than their unipolar counterpart.

